\documentclass[amssymb,amsmath,preprint,prd,nofootinbib]{revtex4-1}
\usepackage[colorlinks]{hyperref}
\usepackage{graphicx}
\usepackage{multirow}

\usepackage{slashed}
\usepackage{gensymb}
\usepackage{orcidlink}
\usepackage{subfigure}

\begin{document}
\title{A model for Axial Non-Standard Interactions of neutrinos with quarks}

\author{S. Abbaslu\, \orcidlink{0000-0003-3567-717X}}
\email{s-abbaslu@ipm.ir}
\author{Y. Farzan\, \orcidlink{0000-0003-0246-5349}}
\email{yasaman@theory.ipm.ac.ir}
\affiliation{School of Physics, Institute for Research in Fundamental Sciences (IPM), PO Box 19395-5531, Tehran, Iran}

%\textcolor{blue}{
%\textcolor{red}{

\begin{abstract}
The neutrino oscillation experiments are setting increasingly strong upper bounds on   the vector  Non-Standard neutrino Interactions 
(NSI) with matter fields. However, the bounds on the axial NSI are more relaxed, raising the hope that studying the neutral current events at an experiment such as DUNE can give a glimpse on new physics. We build a model that gives rise to axial NSI with large couplings leading to observable deviation from the standard prediction at DUNE. The model is based on a $U(1)$ gauge symmetry with a gauge boson of mass $\sim 30$~GeV which can be discovered at the high luminosity LHC. Combining the LHC and DUNE discoveries, we can unravel the axial form of interaction. The cancellation of anomalies of the gauge group suggests new heavy quarks as well as a dark matter candidate. The new quarks mixed with the first generation quarks can also be discovered at the LHC. Moreover, they provide a seesaw mechanism that explains the smallness of the $u$ and $d$ quark masses. The dark matter has an axial coupling to the quarks which makes its discovery via spin dependent direct dark matter search experiments possible.
    
\end{abstract}
\maketitle

\section{Introduction}
After the establishment of the neutrino mass and mixing scheme as a solution for the solar and atmospheric neutrino anomalies, the main goal in this field has been determining the values of still unknown neutrino mass matrix parameters. In particular, to measure the Dirac CP violating phase of the PMNS neutrino mixing matrix, state-of-the-art long baseline experiments such as DUNE are under development. The precision of these experiments will enable us to study subdominant effects that might arise from new physics.  Finding a footprint of new physics in the neutrino experiments is an opportunity that we do not want to miss. A wide class of new physics relevant for neutrino experiments can be described by the following effective potential
\begin{equation}
	\frac{G_F}{\sqrt{2}}[\bar{\nu}_\alpha \gamma^\mu (1-\gamma^5)\nu_\beta][\bar{q}\gamma_\mu(\epsilon_{\alpha\beta}^{Vq}+\epsilon_{\alpha\beta}^{Aq}\gamma^5)q]  \ \ \ \  \  {\rm where}  \ \ \ \ \  q\in \{ u,d\}\ .
\label{eff}\end{equation}
$\epsilon_{\alpha\beta}^{Vq}$ and $\epsilon_{\alpha\beta}^{Aq}
$ are massless parameters quantifying the strength of the coupling. At $\epsilon_{\alpha\beta}^{Vq},\epsilon_{\alpha\beta}^{Aq}\to 0$, we recover the Standard Model (SM) limit.
The impact of the vector coupling $\epsilon_{\alpha\beta}^{Vq}$ on the neutrino experiments has been extensively studied in the literature \cite{Farzan:2017xzy}. As is well-known, this coupling can affect the neutrino propagation in matter as well as the Coherent Elastic $\nu$ Nucleus Scattering (CE$\nu$NS).
From these observations, strong bounds on $\epsilon_{\alpha\beta}^{Vq}$    have been derived
\cite{Coloma:2023ixt}. Also from the model building point of view, $\epsilon_{\alpha\beta}^{Vq}$  has received considerable attention \cite{Babu:2019mfe,Farzan:2016fmy,Farzan:2016wym,Farzan:2015hkd,Farzan:2015doa,Bischer:2018zbd,Forero:2016ghr,Farzan:2019xor}. 
The axial NSI  ($\epsilon^{Aq}$) cannot affect neutrino propagation in matter or CE$\nu$NS but can change the cross section of non-coherent Neutral Current (NC) scattering of neutrinos off nuclei and therefore the number of NC events at the neutrino detectors. In particular, it can affect Deuterium dissociation so by studying the neutral current events at SNO, bounds on $\epsilon^{Au}-\epsilon^{Ad}$
have been derived \cite{Coloma:2023ixt}.  Complementary bounds are derived from high energy neutrino scattering experiment, CHARM \cite{CHARM:1986vuz,CHARM-II:1994dzw,Davidson:2003ha}. 
Combining the SNO results with CHARM, we find the following 90 \% C.L. bounds,
$$ -0.19<\epsilon_{ee}^{Au}-\epsilon_{ee}^{Ad}<0.13 \ \ {\rm and} \ \ -0.13<\epsilon_{e\tau}^{Au}-\epsilon_{e\tau}^{Ad}<0.1\ .$$
Moreover, CHARM implies 
$$ -1.7 <\epsilon_{ee}^{Au}<0.7, 
\ -0.8 <\epsilon_{ee}^{Ad}<0.9 \ {\rm and} \  |\epsilon_{e\tau}^{Aq}|<1 \ .$$
The strongest bounds on the $\mu \mu$ and $\mu \tau$ components come from the NuTeV neutrino scattering experiment
\cite{NuTeV:2001whx}: 
$$ |\epsilon^{Aq}_{\mu\mu}|<0.01 \ \ {\rm and} \ \  |\epsilon^{Aq}_{\mu\tau}|<0.1 \ .$$
The $e\mu $ components are severely constrained by the bound on the lepton flavor violating process $\mu^- {\rm Ti} \to e^- {\rm Ti} $ \cite{Davidson:2003ha}:
$$|\epsilon_{e\mu}^{Aq}|<7.7 \times 10^{-4} \ .$$
The SNO experiment  \cite{Coloma:2023ixt} finds two solutions for the difference of $\epsilon_{\tau\tau}^{Ad}$ and $\epsilon_{\tau\tau}^{Au}$
as $$ -2.1<\epsilon_{\tau\tau}^{Au}-\epsilon_{\tau\tau}^{Ad}<-1.8  \ \ {\rm or} \ \ -0.2<\epsilon_{\tau\tau}^{Au}-\epsilon_{\tau\tau}^{Ad}<0.15 \ . $$
Despite these bounds,   the $\tau \tau$ component  of  the isospin singlet combination, $\epsilon^{Au}+\epsilon^{Ad}$, is still practically unconstrained in the present literature, allowing values even larger than 1 \cite{Coloma:2023ixt}.\footnote{Studying the neutral current at MINOS and MINOS+ can also constrain $\epsilon_{\tau\tau}^{Au}$ and $\epsilon_{\tau\tau}^{Ad}$. Recently, we have set a bound $-0.35<\epsilon_{\tau\tau}^{Au}=\epsilon_{\tau\tau}^{Ad}<0.3$ \cite{us} using the MINOS and MINOS+ data. Moreover in \cite{Abbaslu:2024jzo}, we discuss the possibility of constraining $\epsilon_{\tau\tau}^{Aq}$ by the KamLAND data. }
Ref. \cite{Abbaslu:2023vqk} shows that  studying the neutral current events at the near and far detectors of DUNE can significantly improve the information on the axial NSI. In particular, $\epsilon^{Au}_{\tau\tau}$ and $\epsilon^{Ad}_{\tau\tau}$ can be probed down to $O(0.1)$ \cite{Abbaslu:2023vqk}. %However, the bound on the $\tau \tau$ component is much weaker. The strongest  bound found in the literature is from SNO \cite{}

It will be dramatically interesting if new physics shows itself in the measurement of $\epsilon^{Au}$ and $\epsilon^{Ad}$ at DUNE. The aim of this paper is to build a model that
leads to $\epsilon^{Vu}_{\alpha \beta}, \epsilon^{Vd}_{\alpha \beta}\ll\epsilon^{Au}_{\tau\tau}\sim\epsilon^{Ad}_{\tau\tau}\sim O(1)$. We present a toy model based on a new gauge symmetry with a new gauge boson of mass $\sim 30$~GeV and axial coupling to the first generation quarks. For anomaly cancellation, we need
to add new chiral fermions which will lead to interesting phenomenology, including a seesaw mechanism for the suppression of the $u$ and $d$ masses relative to the electroweak scale as well as a dark matter candidate with a mass of $\sim$ GeV which can be discovered by  spin dependent direct dark matter search experiments.

This paper is organized as follows. In sect. \ref{model}, we present the model and its field content and discuss the production and detection of its dark matter candidate. We also discuss the bounds on the parameter space of the model. In sect. \ref{signal}, we discuss how various new particles introduced in this model can be discovered. In sect. \ref{gener}, we first discuss a generalization of the model to obtain an arbitrary lepton flavor structure for $\epsilon_{\alpha\beta}^{Au}=
\epsilon_{\alpha\beta}^{Ad}$.  We then suggest an alternative UV completion of the model without new colored particles. Our results are summarized in sect. \ref{summary}.
%%%%%%%%%%%%%%%%%%%%%
%%%%%%%%%%%%%%%%%%%%%
\section{The model \label{model}}
%%%%%%%%%%%%%%%%%%%%%%%
%%%%%%%%%%%%%%%%%%%%%%

The effective interaction of the form in Eq. (\ref{eff}) can be obtained by integrating out a massive gauge boson, $Z^\prime_\mu$, with axial couplings to the first generation quarks as well as with couplings to neutrinos. As explained before, $\epsilon_{\tau \tau}^{Aq}$ is the least experimentally constrained element. In this section, we shall introduce a  model for 
$|\epsilon_{\tau \tau}^{Aq}|\sim 1$. Let us introduce a gauge symmetry under which all three standard quark doublets have equal charges
\begin{equation}  Q_1=\left( \begin{matrix} u_L \cr d_L\end{matrix} \right) \to e^{i\alpha } Q_1 \ , \ \ \  Q_2=\left( \begin{matrix} c_L \cr s_L\end{matrix} \right) \to e^{i\alpha } Q_2 \ , \ \ \  {\rm and}  \ \ \ \  Q_3=\left( \begin{matrix} t_L \cr b_L\end{matrix} \right) \to e^{i\alpha } Q_3 \ . 
\end{equation}
Let us now assign opposite $U(1)$ charges to the first generation of the right-handed quarks
$$u_R \to e^{-i\alpha } u_R    \ \ \  {\rm and}  \ \ \ \  d_R \to e^{-i\alpha } d_R \ .$$
Then, the couplings of first generation quarks to $Z'_\mu$ can be written as
\begin{equation}
\begin{split}
&g_{Z^\prime} Z^\prime_\mu \left(\bar{u} \gamma^\mu \frac{1-\gamma_5}{2}u +
\bar{d} \gamma^\mu \frac{1-\gamma_5}{2}d-\bar{u} \gamma^\mu \frac{1+\gamma_5}{2}u  -	\bar{d} \gamma^\mu \frac{1+\gamma_5}{2}d\right)\\&=-g_{Z^\prime} Z^\prime_\mu \left(\bar{u} \gamma^\mu \gamma_5 u +
\bar{d} \gamma^\mu \gamma_5 d\right)\ .
\end{split}
\end{equation}
We have therefore obtained the axial form of coupling as we desired. Notice that our model respects the isospin symmetry. To cancel the $U(1)-SU(2)-SU(2)
$  anomaly, we should assign $U(1)$ charges to lepton doublets, too. Since we want only $\nu_\tau$ to obtain effective coupling, we should assign a charge of $-9$ to $L_3$:
\begin{equation}  L_\tau=\left( \begin{matrix} \nu_\tau \cr \tau_L\end{matrix} \right) \to e^{-9 i\alpha } L_\tau   \ \ \ \ {\rm under \ }U(1)\ .
\end{equation}

We then obtain
\begin{equation} \label{epp}\epsilon_{\tau \tau}^{Au}=\epsilon_{\tau \tau}^{Ad}=\frac{9\sqrt{2} g_{Z^\prime}^2}{m_{Z'}^2 G_F}. 
\end{equation}
As mentioned above, with this $U(1)$ charge assignments to the standard left-handed fermions, the $U(1)-SU(2)-SU(2)$ anomaly will cancel out. However,  we should also ensure the cancellation of the $U(1)-SU(3)-SU(3)$  anomaly.

Before introducing our mechanism for the cancellation of the $U(1)-SU(3)-SU(3)$ anomaly, let us discuss the $U(1)$ charges of the right-handed fermions as well as that of the Higgs. We denote the $U(1)$ charges of $\tau_R$ $c_R$, $s_R$, $t_R$ and $b_R$ respectively by $\alpha_\tau$, $\alpha_c$, $\alpha_s$, $\alpha_t$ and $\alpha_b$.  The measured values of the Yukawa couplings of the SM Higgs to $\tau$, $b$ and $t$ are in remarkable agreement with the SM prediction \cite{ATLAS:2024fkg}. Thus, it is established that the mass generation mechanism for the third generation fermions is indeed the SM Higgs mechanism. 
Denoting the $U(1)$ charge of the Higgs by $\alpha_H$, this implies $\alpha_\tau={-9}-\alpha_H$, $\alpha_b=1-\alpha_H$ and $\alpha_t+\alpha_b=2$.  Throughout this paper, we shall set $\alpha_H=0$ so we do not need to worry about the tree level decay modes such as $H\to Z'Z'$. The Yukawa couplings of the second generation quarks are less constrained by observation \cite{ATLAS:2024fkg} so it is imaginable that they acquire mass through couplings to another Higgs doublet with a $U(1)$ charge of $\alpha_{H'}$. For example, if we introduce a new Higgs that couples exclusively to $s_R$ and $Q_2$, the $U(1)$ charge cancelation implies $\alpha_s+\alpha_{H'}=\alpha$ so, with an arbitrary value of $\alpha_{H'}$, the mass generation for the $s$ quark does not dictate the equality of $\alpha_s$ and $\alpha$.  Taking
$\alpha_s=8-\alpha_c-\alpha_t-\alpha_b$, the $U(1)-SU(3)-SU(3)$ anomaly cancels out without  needing new quarks. However, the mass generation for the first and second generations requires four new Higgs doublets with non-trivial $U(1)$ charges.
In sect. \ref{gener}, we shall discuss the variant of our model without new quarks.

Another way to cancel the $U(1)-SU(3)-SU(3)$ anomaly is to add new colored particles charged under the new $U(1)$.
Adding new colored particles makes the model more elegant and rich from the phenomenological point of view. In the following, we shall complete our model by adding new quarks.  
Let us introduce new quarks, $T_L$, $T_R$, $B_L$ and $B_R$ which are singlets of standard $SU(2)$ and are in the fundamental representation (triplet) of the $SU(3)$ gauge group. The  electric charges of $B$ ({\it i.e.,} $B_L$ and $B_R$) and  $T$ ({\it i.e.,} $T_L$ and $T_R$) are respectively equal to $-1/3$ and $2/3$ as SM quarks. The field content of our model as well as their quantum numbers are shown in Table \ref{tab:charges2}.  The first and second generations of the leptons as well as the SM Higgs are taken to be neutral under the new $U(1)$ and are not shown in the table \ref{tab:charges2}.  Notice that $s_R$, $c_R$, $b_R$, and $t_R$ with charges opposite to those of $u_R$ and $d_R$ are taken to be mass eigenstates (rather than flavor eigenstates) to avoid Flavor Changing Neutral Currents (FCNC). We shall discuss more on FCNC  in sect. \ref{signal}.

\begingroup

\setlength{\tabcolsep}{7pt} % Default value: 6pt
\renewcommand{\arraystretch}{1.6} % Default value: 1

%\begin{table}
	%\centering
	%\caption{Particles and their relevant charges under the gauge symmetry group. Here, T, D, and S stand for Triplet, Doublet, and Singlet respectively. The first and second lepton generations are not charged under new $\rm U(1)$ gauge, but have the same properties under the $\rm SU(3)$, $\rm SU(2)$, and $\rm U_{Y}(1)$ as the third lepton generation.}
%\begin{tabular}{ |c|c|c|c|c|c|c|c| c|c|c|c|c|c|  }
%	\hline
%	\backslashbox{Gauge group}{Particles} & $Q_i$ &$t_R,c_R$&$b_R,s_R$&$u_R$ &$d_R$&$L$&$\tau_R$&$Q_4$&$Q_{4_R}$&$\chi_L$&$\phi$&$\phi^{\prime}$\\
	%\hline
	%$\rm SU(3)$  & T & T & T & T & T & S & S & T & T & S & S & S\\
	%\hline
	%$\rm SU(2)$  & D & S & S & S & S & D & S & D & D & S & S & S\\
	%\hline
	%$\rm U_{Y}(1)$ & $\frac{1}{6}$ & $\frac{2}{3}$ & $-\frac{1}{3}$ & $\frac{2}{3}$ & $-\frac{1}{3}$ & $-\frac{1}{2}$ & -1 & $\frac{1}{6}$ & $\frac{1}{6}$ & $0$ & $0$  & $0$ \\
	%\hline
	%$\rm U(1)$    & $1$ & $1$ & $1$ & $-1$ & $-1$ & $-3$ & $-3$ & $-1$ & $1$ & $3$ & $2$ & $-6$\\

%\hline
%\end{tabular}\label{tab:charges}
%\end{table}

\begin{table}[htb!]
	\caption{Particles and their relevant quantum number under the gauge symmetry group. Here, T, D, and S stand for Triplet, Doublet, and Singlet respectively. The first and second generation leptons are not charged under the new $\rm U(1)$ gauge and are not shown here.  } %title of the table
\centering % centering table
\begin{tabular}{ccccccccccccccc} % creating eight columns
\hline\hline %inserting double-line
 &\multicolumn{9}{c}{Particles} \\ [0.9ex]
\hline % inserts single-line
& $Q_i$ & $t_R,c_R$ & $b_R,s_R$ & $u_R$ & $d_R$ & $L_\tau$ & $\tau_R$ & $T_R$ & $T_L$& $B_R$ & $B_L$& $\chi_L$ & $\phi$ & $\phi^{\prime}$\\
\hline
$\rm SU(3)$  & T & T & T & T & T & S & S & T & T&T & T& S & S & S\\
\hline
$\rm SU(2)$  & D & S & S & S & S & D & S & S& S &S & S& S & S & S\\
\hline
$\rm U_{Y}(1)$ & $\frac{1}{6}$ & $\frac{2}{3}$ & $-\frac{1}{3}$ & $\frac{2}{3}$ & $-\frac{1}{3}$ & $-\frac{1}{2}$ & -1 & $\frac{2}{3}$ & $\frac{2}{3}$ & $-\frac{1}{3}$ & $-\frac{1}{3}$ & $0$ & $0$  & $0$ \\
\hline
$\rm U(1)$    & $1$ & $1$ & $1$ & $-1$ & $-1$ & $-9$ & $-9$ & $1$ & $-1$  & $1$ &  $-1$ & $9$ & $-2$ & $-18$\\
\hline % inserts single-line
\end{tabular}
\label{tab:charges2}
\end{table}

\endgroup

%\begingroup

%\setlength{\tabcolsep}{10pt} % Default value: 6pt
%\renewcommand{\arraystretch}{1.6} % Default value: 1
%\begin{table}[h]
%	\caption{Particles and their relevant charges under the gauge symmetry group. Here, T, D, and S stand for Triplet, Doublet, and Singlet respectively. The first and second lepton generations are not charged under new $\rm U(1)$ gauge, but have the same properties under the $\rm SU(3)$, $\rm SU(2)$, and $\rm U_{Y}(1)$ as the third lepton generation.  } %title of the table
	%\centering % centering table
	%\begin{tabular}{ccccccccccccc} % creating eight columns
	%	\hline\hline %inserting double-line
		%  &\multicolumn{9}{c}{Particles} \\ [0.9ex]
	%	\hline % inserts single-line
	%		& $Q_i$ &$t_R,c_R$&$b_R,s_R$&$u_R$ &$d_R$&$L$&$\tau_R$&$Q_4$&$Q_{4_R}$&$\chi_L$&$\phi$&$\phi^{\prime}$\\
	%	\hline
 %$\rm SU(3)$  & T & T & T & T & T & S & S & T & T & S & S & S\\
%%		\hline
%		$\rm SU(2)$  & D & S & S & S & S & D & S & D & D & S & S & S\\
%		\hline
%		$\rm U_{Y}(1)$ & $\frac{1}{6}$ & $\frac{2}{3}$ & $-\frac{1}{3}$ & $\frac{2}{3}$ & $-\frac{1}{3}$ & $-\frac{1}{2}$ & -1 & $\frac{1}{6}$ & $\frac{1}{6}$ & $0$ & $0$  & $0$ \\
%		\hline
%		$\rm U(1)$    & $1$ & $1$ & $1$ & $-1$ & $-1$ & $-3$ & $-3$ & $-1$ & $1$ & $3$ & $2$ & $-6$\\
%		\hline % inserts single-line
%	\end{tabular}
%	\label{tab:charges2}
%\end{table}

%\endgroup

With the field content shown in table \ref{tab:charges2}, all gauge anomalies cancel out. The left-handed  $\chi_L$ is added to cancel the remaining $U(1)-U(1)-U(1)$ anomaly. Being neutral, $\chi_L$ can be a suitable DM candidate with a dazzling prospect of direct detection as we discuss below.  Notice that with this charge assignment, the $u$ and $d$ quarks cannot obtain mass via Yukawa coupling to the SM Higgs. In our model, their masses come from mixing with the fourth generation quarks.
Indeed, this model also explains the hierarchy between the first and third generation quarks as a bonus in an elegant fashion. 
Notice that we could put $(B_L \ T_L)$ as well as $(B_R \ T_R)$
in the $SU(2)$ electroweak doublets but the anomaly cancellation would then require smaller $U(1)$ charge for $L_\tau$, leading to smaller $\epsilon_{\tau \tau}^{A u}=\epsilon_{\tau \tau}^{A d}$.

The terms that lead to mass generation for the new fermions are
 \begin{equation} 
  \begin{split}
 &\lambda_{B} \phi \bar{B}_{L} B_R +\lambda_{T}\bar{T}_{L}T_R \phi
 +\frac{1}{2}\lambda_\chi \phi^\prime \chi_L^T c\chi_L+ M_{Bd} \bar{B}_L d_R
 	+M_{Tu}\bar{T}_L u_{R} 	+\lambda_{Bd} \bar{B}_R H^{\dagger}Q_1 \\&+   \lambda_{Tu} \epsilon_{ij} H_i \bar{T}_R(Q_1)_j+{\rm H.c.} \ \  {\rm where} \ \ c=\left( \begin{matrix} 0 & 1\cr -1 & 0 \end{matrix}\right) . \label{Yukawa}
  \end{split}
 	\end{equation}
  Like the matrix $c$, $\epsilon_{ij}$ is also an anti-symmetric matrix with off-diagonal elements equal to $\pm 1$. While the $\epsilon$ matrix acts on the $SU(2)$ indices, the $c$ matrix acts on the spinorial indices of the Weyl fermions.
  We could add interactions to the second and third generation quarks as $\bar{B}_R H^\dagger Q_{2,3}$ and $ \epsilon_{ij} H_i \bar{T}_R(Q_{2,3})_j$. For simplicity, we forbid these terms with an approximate global $U(1)$ symmetry under which only the second and third generation quarks are charged. Spontaneous symmetry breaking leads to 
 a Majorana mass for   $\chi_L$
  \begin{equation}  \frac{1}{2}m_\chi \chi_L^T c\chi_L \ \ \ {\rm where} \ \ \ 
  	m_\chi= \lambda_\chi \langle \phi'\rangle  \ . \ \ 
  \ \end{equation}
Notice that $\chi_L^T c=\bar{\chi}_L^c$. For a detailed discussion of the CP-conjugation of the Majorana fields, the readers may consult Ref. \cite{Fujikawa:2024oyf}.
Moreover, vacuum expectation values of $\phi$ and $\phi'$ lead to a mass for the gauge boson 
\begin{equation}
 \ \ \label{mZ}m_{Z'}=\sqrt{2}g_{Z^\prime}\left(4  \langle \phi \rangle^2 +324  \langle \phi'\rangle^2\right)^{1/2} \ .\end{equation}

 After the $U(1)$ and $SU(2) \times U_Y(1)$ spontaneous symmetry breaking, 
   the first and fourth generation quarks obtain mass and mix as
 \begin{equation}
	(\bar{u}_R  \  \bar{T}_R) \left(\begin{matrix} 0 & M_{Tu}^{\star} \cr \lambda_{Tu} \langle H\rangle & \lambda_T \langle \phi\rangle \end{matrix}\right)\left( \begin{matrix} u_L \cr T_L \end{matrix}\right) +(\bar{d}_R  \  \bar{B}_R) \left(\begin{matrix} 0 & M_{Bd}^{\star} \cr \lambda_{Bd} \langle H\rangle &\lambda_{B}\langle \phi\rangle \end{matrix}\right)\left( \begin{matrix} d_L \cr B_L. \end{matrix}\right) .\end{equation} 
Taking  $M_{Tu}^{\star}$, $\lambda_{Tu} \langle H\rangle \ll \lambda_T \langle \phi\rangle $ and $M_{Bd}^{\star}$, $\lambda_{Bd} \langle H\rangle \ll \lambda_B \langle \phi\rangle $, the heavier mass eigenstates ({\it i.e.,} the masses of the fourth generation quarks $T$ and $B$) will be equal to 
$$ \ \  M_B=\lambda_B \langle\phi \rangle , \ \ {\rm and} \ \  M_T=\lambda_T \langle\phi \rangle\ .$$
The lighter mass eigenvalues, which are the masses of the mass eigenstates $\tilde{u}$ and $\tilde{d}$, are respectively given by
\begin{equation} m_u=\frac{\lambda_{Tu}\langle H\rangle M_{Tu}^{\star} }{\lambda_{T} \langle \phi\rangle}=\frac{\lambda_{Tu}\langle H\rangle M_{Tu}^{\star} }{M_T} \ \ {\rm and } \ \  m_d=\frac{\lambda_{Bd}\langle H\rangle M_{Bd}^{\star} }{\lambda_{B} \langle \phi\rangle}=\frac{\lambda_{Bd}\langle H\rangle M_{Bd}^{\star} }{M_B}. \label{light-quark-masses}
\end{equation}
 Relations in Eq.~(\ref{light-quark-masses}) are written at the energy scale around $M_B$ and $M_T$. To compare with the so-called ``current quark masses" reported in PDG \cite{ParticleDataGroup:2024cfk} at the energy scale $\sim 2$~GeV, we should take into account the running of the quark masses with energy which results in an enhancement of order of one \cite{Chetyrkin:1999pq}. Notice that the running of quark masses from the mass scale of $T$ and $B$ down to 2~GeV is similar to SM because the new quarks decouple below their mass scale. The coupling of the new gauge interaction is much smaller than the QCD coupling and its contributions to the running of the masses are therefore negligible. In our model, there is no upper bound on the mass parameters $M_{Bd}$ and $M_{Tu}$. However, their smallness relative to other mass scales of the model is ``technically natural." That is at the limit that $M_{Bd},M_{Tu} \to 0$, a global symmetry under which $\phi\to e^{i\alpha}\phi$,
$T_L\to e^{i\alpha}T_L$ and
$B_L\to e^{i\alpha}B_L$
is restored, satisfying the famous 't Hooft criterion for naturalness \cite{Dine:2015xga}. Notice that $M_{Bd}$ and $M_{Tu}$ are dimensionful parameters so the breaking of the global symmetry is soft which implies that the radiative corrections to $M_{Bd}$ and $M_{Tu}$ are proportional to themselves. Below the mass scale of $T$ and $B$, these new quarks decouple so the running of $M_{Bd}$ and $M_{Tu}$ below this scale is of no relevance.
With $M_T\gg M_{Tu}, \lambda_{Tu} \langle H\rangle$ and $M_B\gg M_{Bd}, \lambda_{Bd} \langle H\rangle$, there will be a seesaw mechanism that explains the smallness of the $u$ and $d$ masses relative to the electroweak scale.
The mixing angles are as follows
\begin{eqnarray} 
{\rm Mixing \ between \ }u_R\  {\rm and} \  T_R= \frac{M_{Tu}^{\star}}{M_T}&,&   
{\rm Mixing \ between \ }u_L\  {\rm and} \  T_L=\sin \alpha=\frac{\lambda_{Tu} \langle H\rangle}{M_T}, \cr  
\cr
{\rm Mixing \ between \ }d_R\  {\rm and} \  B_R=\frac{M_{Bd}^{\star}}{M_B} &,&   {\rm Mixing \ between \ } d_L
 {\rm and} \  B_L=\sin \beta=\frac{\lambda_{Bd} \langle H\rangle}{M_B}.\cr
\end{eqnarray}
Through the mixing, the mass eigenstate $\tilde{T}$ and $\tilde{B}$ can decay to the first generation quarks via decay modes such as $\tilde{T}\to W+\tilde{d}$ and $\tilde{B}\to W+\tilde{u}$. Since the $U(1)$ charges of $T_{R}$ and $B_R$ (of $T_{L}$ and $B_L$) are opposite to those of $u_R$ and $d_R$ (of $Q_1$),  in the mass basis, $Z'$ can have flavor changing couplings to the first and fourth generation quarks with a coupling given by  $2g_{Z^\prime}$ times the relevant mixing. Both decays to $W$ and to $Z'$ are suppressed by the mixing but since we expect $g_{Z^\prime}$ to be much smaller than the $SU(2)$ coupling, the decay to the $W$ boson will dominate. In our model, since the fourth generation left-handed quarks are $SU(2)$ singlets, their electroweak couplings are different from those of the first generation. As a result, the $Z$ boson can also have flavor changing coupling between left-handed first and fourth generation quarks leading to $T \to Zu$ and $B\to Zd$. The rates of these decay modes are comparable to the charged current ones.
The bound on the masses of new quarks, which mainly decay into $W$ and first generation quarks, is \cite{ATLAS:2024zlo}
\begin{equation}
M_T, M_B>1530~{\rm GeV}\ .
\end{equation}
%\textcolor{blue}{ The perturbativity condition is satisfied at the tree level when the  Yukawa coupling is $\lambda<4\pi$. To avoid the problem of radiative corrections in the presence of large fourth-generation Yukawa couplings, it is common to assume that the fourth-generation model acts as an effective theory that breaks down at a scale above the TeV scale \cite{Bellantoni:2012ag, Hung:2009ia, Kribs:2007nz}. In the 2HDM model, it is assumed that the effective potential can be stabilized by new effects near the few TeV scale, and that the effective potentials remain valid below this scale \cite{Bellantoni:2012ag}.}
%\textcolor{cyan}{For example, it is shown that in the context of the MSSM with a fourth generation of chiral matter, introducing one extra vector-like state at the TeV scale allows perturbativity all the way up to the GUT scale \cite{Murdock:2008rx}.....}  

\begin{figure}[] 
\centering
\includegraphics[width=1.\textwidth]{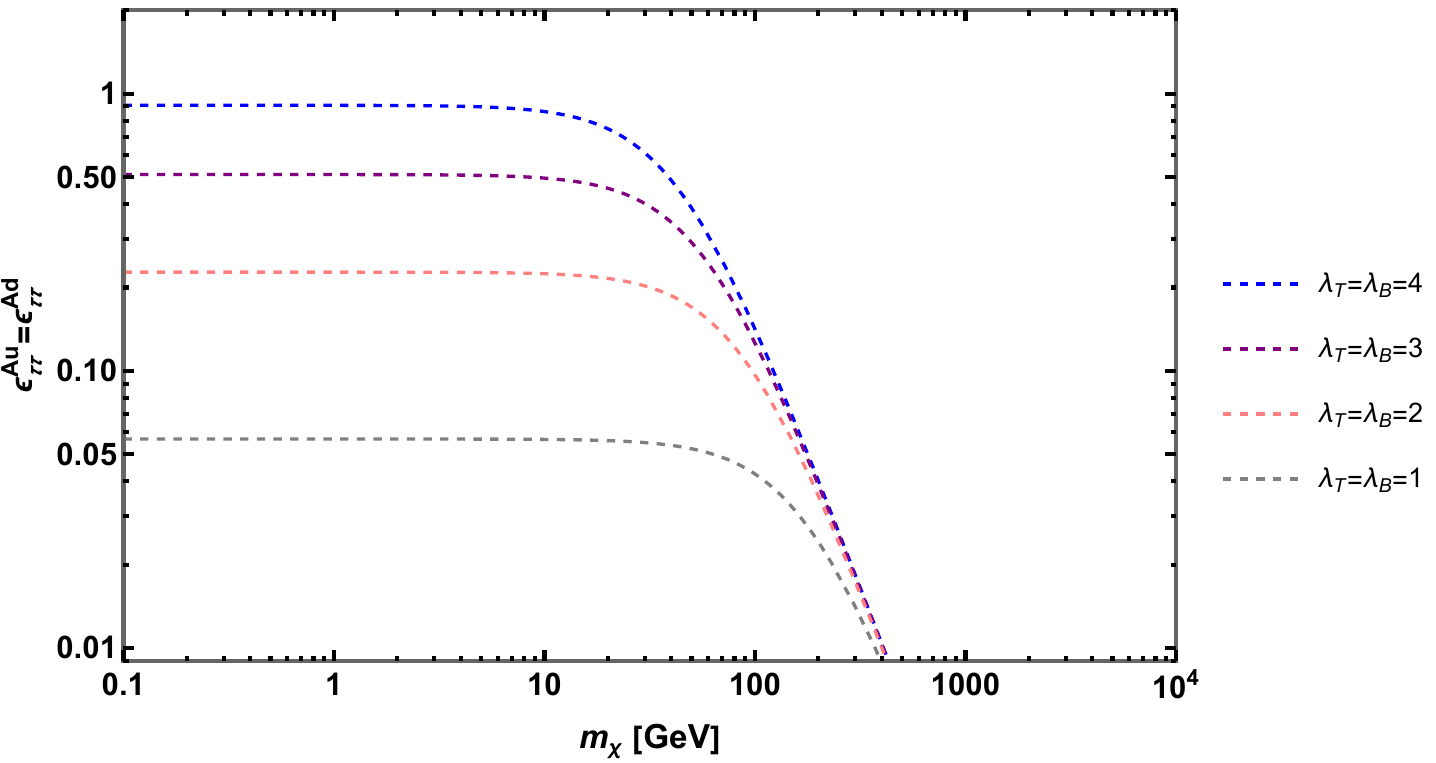}
\vspace{-1.5em}
\caption{ $\epsilon_{\tau\tau}^{Au}$ and $\epsilon_{\tau\tau}^{Ad}$ versus $m_\chi$ for $\lambda_{\chi}=1$ and four different values: $\lambda_B=\lambda_T=4$ (dashed blue), $\lambda_B=\lambda_T=3$ (dashed purple), $\lambda_B=\lambda_T=2$ (dashed pink), and $\lambda_B=\lambda_T=1$ (dashed gray). We have fixed $M_T$ and $M_B$ to 1550 GeV.} 
\label{fig:epsilon_plots}
\end{figure}

Taking $\lambda_T$ and $\lambda_B$ in the perturbative range $(\lambda_T ,\lambda_B<4)$, Eqs. (\ref{epp},\ref{mZ}) yield 
\begin{equation}
	\epsilon_{\tau \tau}^{Au}=\epsilon_{\tau \tau}^{Ad}<0.9\ .\label{bound}
\end{equation} 
As a result, our model can lead to  $	\epsilon_{\tau \tau}^{Au}=\epsilon_{\tau \tau}^{Ad}> 0.1$ which can be probed by DUNE \cite{Abbaslu:2023vqk}.
 Notice that with large $\lambda_T$ and $\lambda_B$, the radiative corrections above the $T$ and $B$ masses can be significant; however, as long as the loop suppression factor, $N_c \lambda_T^2/(16 \pi^2)$ or $N_c \lambda_B^2/(16 \pi^2)$ (with $N_c=$number of colors=3) is smaller than 1, the model stays in the perturbative regime \cite{Ishiwata:2011ny,Allwicher:2021rtd}.
Figure~\ref{fig:epsilon_plots} shows $\epsilon_{\tau\tau}^{Au}=\epsilon_{\tau\tau}^{Ad}$ versus $m_\chi$ for different values of  the Yukawa couplings, fixing $M_T$
and $M_B$ to 1550~GeV. For $m_\chi<10$~GeV, $\langle \phi\rangle \ll M_T/\lambda_T, M_B/\lambda_B$ so the NSI coupling becomes independent of $m_\chi$. 
Notice that saturating the bound in Eq.~(\ref{bound}) requires $9\langle \phi'\rangle \ll \langle \phi\rangle$ (and therefore $m_\chi \ll M_T , M_B$). As we shall see below, this is the range that we are interested in. Taking $M_{Z'}> 30$~GeV, we can safely use the four fermion effective interaction for neutrino scattering in experiments such as DUNE. Such large values of $m_{Z'}$ requires $g_{Z^\prime}>0.02.$  On the other hand, to avoid the present bound \cite{CMS:2019xai}, $g_{Z^\prime}$ should be smaller than 0.2. 

Inserting $m_u\sim m_d\sim 5 $~MeV in Eq. \ref{light-quark-masses}, we find that $\lvert\lambda_{Tu} M_{Tu}^{\star}\lvert\sim 0.045~{\rm GeV} (M_T/1550~{\rm GeV})$ and $\lvert\lambda_{Bd} M_{Bd}^{\star}\lvert\sim 0.045~{\rm GeV} (M_B/1550~{\rm GeV})$. $Q_1$ and $T_R$, being heavier than the Higgs, cannot appear in the decay modes of $H$. Moreover, we can take $\lambda_{Tu},\lambda_{Bd}\ll \lambda_t$ so that the radiative correction due to these couplings to Higgs observables such as the Higgs mass or the Higgs decay modes will be negligible compared to those from the SM top Yukawa coupling.

The Lagrangian can also include the following quartic terms \begin{equation}\lambda_H|H|^4+\lambda_\phi|\phi|^4+\lambda_{\phi^\prime}|\phi^\prime|^4+\lambda_{\phi \phi^\prime}|\phi|^2|\phi^\prime|^2+\lambda_{H \phi}|H|^2|\phi|^2+\lambda_{H \phi^\prime}|H|^2|\phi^\prime|^2\ .\end{equation} The ``unbounded from below condition" implies $\lambda_H,\lambda_\phi,\lambda_{\phi^\prime}>0$, $\lambda_{\phi \phi^\prime}>-2\sqrt{\lambda_{\phi}\lambda_{\phi^\prime}}$, $\lambda_{H \phi}>-2\sqrt{\lambda_{\phi }\lambda_{H}}$ and  $\lambda_{H \phi^\prime}>-2\sqrt{\lambda_{\phi^\prime }\lambda_{H}}$. The couplings $\lambda_H$, $\lambda_\phi$ and $\lambda_{\phi^\prime}$ should be nonzero to induce vacuum expectation value for the corresponding scalar fields. However, $\lambda_{\phi \phi^\prime}$, $\lambda_{H\phi }$ and $\lambda_{H \phi^\prime}$ can be arbitrarily small or even zero without affecting the results of our paper. Indeed from various considerations, there are bounds on these couplings: The stability of masses of the scalars requires $\lambda_{\phi \phi^\prime}\ll m_{\phi^\prime}^2/\langle \phi\rangle^2$, $\lambda_{H \phi^\prime}\ll m_{\phi^\prime}^2/\langle H\rangle^2$ and $\lambda_{H \phi}\ll m_{H}^2/\langle \phi\rangle^2$. Moreover, to make $Br(H\to \phi^\prime  \phi^\prime)$ smaller than $Br(H \to b\bar{b})$, it is required $\lambda_{H\phi^\prime}\ll m_b/\langle H\rangle=0.02$. Throughout our analysis, we set $\lambda_{H\phi^\prime}=\lambda_{\phi\phi^\prime}=\lambda_{H\phi}=0$. Then, we simply have \begin{equation} 
M_H=2\sqrt{\lambda_H}\langle H\rangle, \ m_\phi=2\sqrt{\lambda_\phi}\langle \phi\rangle \ \ {\rm and}\ \ m_{\phi^\prime}=2\sqrt{\lambda_{\phi^\prime}}\langle \phi^\prime\rangle \ . \label{scalar-masses}
  \end{equation}
	
	Let us now discuss whether $\chi_L$ can play the role of Dark Matter (DM). To cancel the $U(1)-U(1)-U(1)$ anomaly, we had to add a SM singlet chiral field, $\chi_L$. In principle, our gauge group allows a coupling of form $\bar{\chi_L^c}\epsilon_{ij} H_i(L_\tau)_j$ which leads to the decay of $\chi_L$ to $\nu_\tau$ plus muon or $s$-quarks pairs. Imposing a $Z_2 $ symmetry under which only $\chi_L$ is odd this term will be forbidden and $\chi_L$ can be stable. Integrating out the intermediate $Z'$,
the effective coupling between $\chi$ and the first generation quarks can be written as 
 \begin{equation}
 	\frac{9 g_{Z^\prime}^2}{m_{Z'}^2}\left(\bar{\chi}\gamma^\mu (\frac{1-\gamma_5}{2})\chi \right)\left(\bar{q} \gamma_\mu \gamma_5 q\right).
 \end{equation}
 This effective interaction leads to spin-dependent interaction between dark matter and the nucleons (proton and neutron)  with the following cross section 
\cite{Cirelli:2024ssz}:
%\begin{equation} 	\sigma_{SD}^N\simeq 7 \left( \frac{m_N m_\chi}{m_N +m_\chi}\right)^2\frac{9g_{Z^\prime}^4}{m_{Z'}^4}=7   \left( \frac{m_N m_\chi}{m_N +m_\chi}\right)^2 \frac{9}{\left(  (2M_4/\lambda_4)^2+(18 m_\chi/\lambda_\chi)^2\right)^2}
	\ .
 %\end{equation}

\begin{equation} 
	\sigma_{SD}^N\simeq 20.97 \left( \frac{m_N m_\chi}{m_N +m_\chi}\right)^2\frac{g_{Z^\prime}^4}{m_{Z'}^4}=5.24  \left( \frac{m_N m_\chi}{m_N +m_\chi}\right)^2 \frac{1}{\Big(  (2M_T/\lambda_T)^2+(18 m_\chi/\lambda_\chi)^2\Big)^2}
	\ .
\end{equation}

\begin{figure}[] 
    \centering
    \subfigure[]{
    \includegraphics[width=0.9\textwidth]{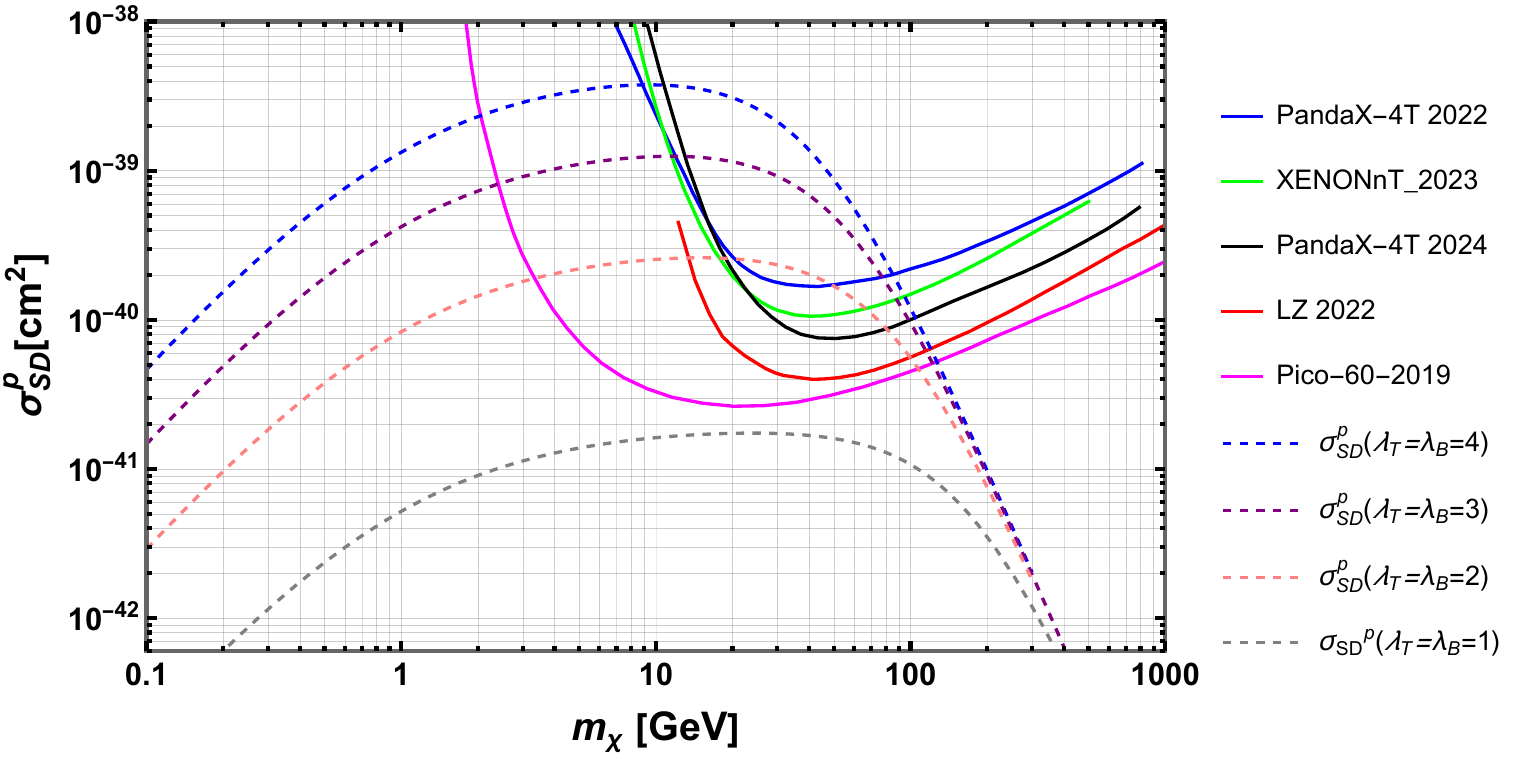}}
    \hspace{0.0mm}
    \subfigure[]{
    \includegraphics[width=0.9\textwidth]{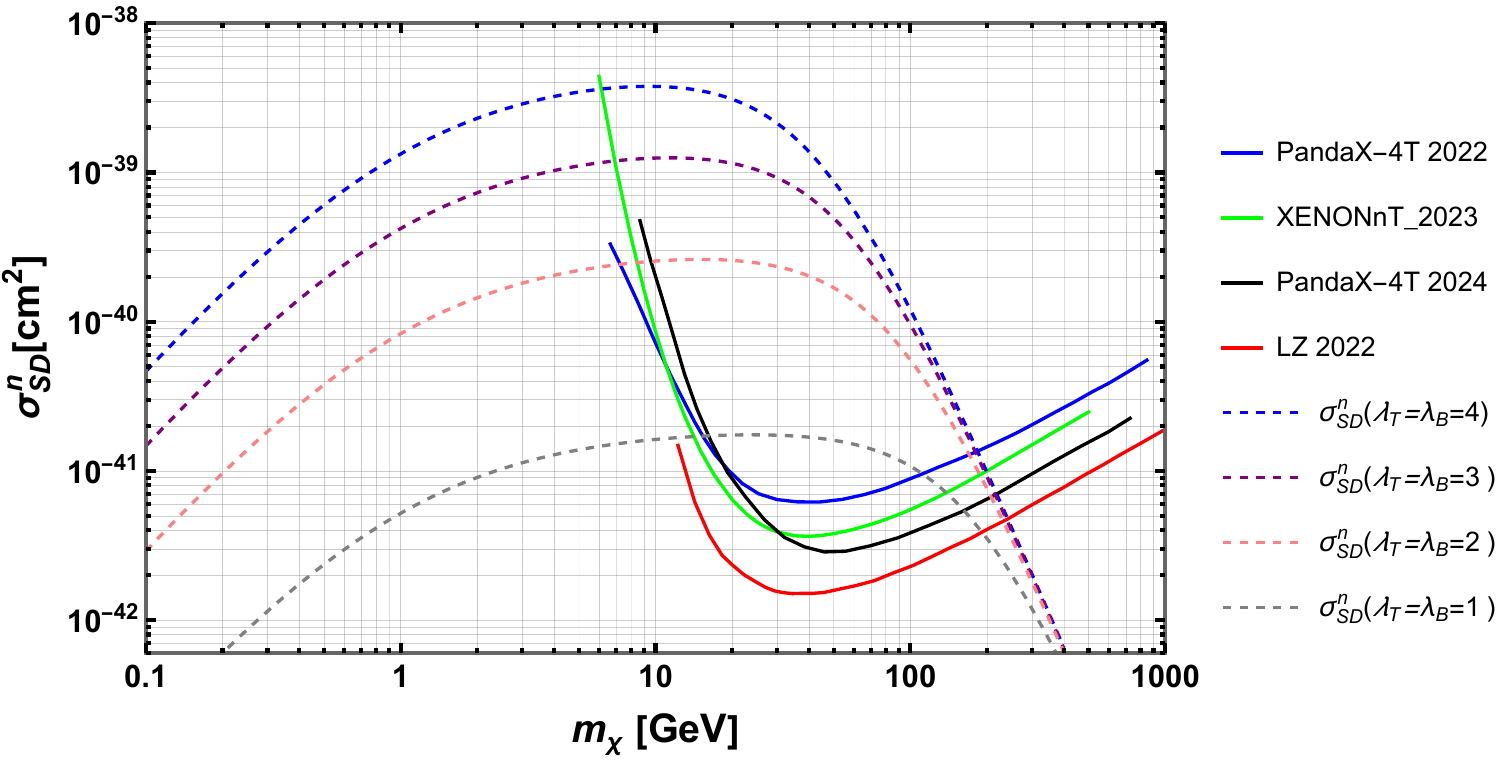}}
    \vspace{0.em}
    \caption{Spin-dependent DM scattering cross section off protons (upper panel)  and off neutrons (lower panel) versus DM mass. The solid curves are the present bounds from spin dependent direct dark matter search experiments and the dashed curves are the predictions of our model. The blue solid line shows PandaX-4T 2022 \cite{PandaX:2022xas}, the green solid line depicts XENONnT 2023 \cite{XENON:2023cxc}, the black solid line represents PandaX-4T 2024 \cite{PANDA-X:2024dlo}, the red solid line shows LZ 2022 \cite{LZ:2022lsv} and finally, the magenta solid line represents PICO-60 \cite{PICO:2019vsc}. The spin dependent cross section $\sigma^{N}_{SD}$ is plotted for $\lambda_{\chi}=1$,  $M_T=M_B= 1550$ GeV and four different values of the quark couplings: $\lambda_B=\lambda_T=4$ (dashed blue), $\lambda_B=\lambda_T=3$ (dashed purple), $\lambda_B=\lambda_T=2$ (dashed pink) and $\lambda_B=\lambda_T=1$ (dashed gray).}  
    \label{fig:both_plots}
\end{figure}

 Notice that our model is isospin invariant, predicting equal cross sections for scattering off the neutron and the proton.
Figure \ref{fig:both_plots} shows the prediction of our model for the scattering cross section for different values of $\lambda_T =\lambda_B$ along with the bounds from the spin dependent direct detection. As seen from these figures, the direct dark matter detection bounds from the PICO experiment \cite{PICO:2019vsc}  forces $\chi$ to be lighter than a few GeV:
\begin{equation} m_\chi<2~{\rm GeV} \ \ {\rm for } \ \lambda_T =\lambda_B=4\ . \end{equation}
Notice that $\chi$ can still be much heavier than few keV which is the lower bound from the structure formation consideration \cite{Cirelli:2024ssz}.  Taking $m_{Z'}>30$~GeV, we can therefore use the four Fermi effective formalism to compute the cross section  of the annihilation of dark matter pair to the standard model fermions \cite{Beltran:2008xg}.
The cross section of the annihilation to the SM fermions for $\langle \phi \rangle >400$~GeV ({\it i.e.,} $M_T, M_B >1600$~GeV) will be much smaller than 1~pb, satisfying the bounds from the indirect dark matter searches \cite{Beltran:2008xg}. %However, if the reheating temperature is higher than the mass of $\chi$ and $Z'$, they will come into thermal equilibrium with the plasma in the early universe.

The aim of this paper is to build, for the first time, an underlying model for $\epsilon^{Au}_{\tau\tau}=
\epsilon^{Ad}_{\tau\tau}\sim 1$. We realized that the model includes a colorless electroweak singlet which can be a suitable dark matter candidate.
 In the literature, a myriad of production mechanisms for dark matter has been suggested. Below, we briefly mention a few possibilities to be incorporated into the model. We emphasize that these mechanisms do not affect the phenomenology of neutrino interaction with matter which is the focus of the present paper. Being only peripheral to our discussion, we shall not go into details and will keep the focus on $\epsilon^{Au}=
\epsilon^{Ad}$.\begin{itemize} \item  i) {\it Low reheating temperature cosmologies:} As shown in Ref. \cite{Hannestad:2004px}, the reheating temperature can be as low as 4~MeV. If the reheating temperature is smaller than the mass of DM, the initial DM number density would be Boltzmann suppressed. Recently a new class of DM production mechanisms has been suggested that is based on low reheating temperatures \cite{Gan:2023jbs,Boddy:2024vgt}. This mechanism can be readily incorporated into our model without affecting our main results.
\item {\it ii) Freeze-out scenario:}
As is well-known in order to obtain the observed DM abundance via the freeze-out scenario, the annihilation cross section should be of the order of 1~pb. A detailed study of the freeze-out scenario for DM of GeV mass scale is carried out in \cite{Steigman:2012nb}.
 To obtain $\sigma_{ann}\sim 1$~pb, we need to turn on new interaction(s). The $\phi'$ particle can be a $s$-channel mediator for this annihilation. For example, if $\phi'$ mixes with the neutral component of a $SU(2)$ triplet with the same $U(1)$ charge, we can obtain annihilation into neutrinos, $\chi \chi \to \nu_\tau \nu_\tau$, which is less constrained by indirect dark matter searches than annihilation to quarks or tau. 
Another option is the annihilation of the DM pair to a pair of light dark sector particles via $s$-channel $\phi'$ exchange. Further elaboration is beyond the scope of the present paper.\end{itemize} %Since this discussion is not an integral part of our model, we postpone describing the annihilation and freeze-out mechanism to the appendix.
%This can take place via a quartic coupling of form $\lambda_{a\phi'}|\phi'|^2  a^2$. The annihilation cross section will be given by 
%\begin{equation}
%\sigma v = \frac{\lambda_{a\phi'}^2m_\chi^2}{128 \pi(m_{\phi'}^2-4m_\chi^2)^2}\sqrt{1-\frac{m_a^2}{m_\chi^2}} \Big(1+\mathcal{O}(v^2)\Big), 
%\end{equation}
 %which can be equal to 1~pb for $m_{\phi'}\sqrt{1-4m_\chi^2/m_{\phi'}^2}=31~{\rm GeV} \sqrt{\lambda_{a\phi'} m_\chi/{\rm GeV}}$. If $m_\chi \sim$GeV, the freeze-out will take place around the time of QCD phase transition so the density of $a$ at the neutrino decoupling era will be suppressed and it will contribute less than 1 to the relativistic degrees of freedom ($\Delta N_{eff}$) at the neutrino decoupling era and later. At temperatures below $m_\chi/20$, the $a$ particles will decouple from the plasma and their density will be entropy suppressed. Taking the $a$ mass smaller than $\sim 100$~eV, their contribution to the matter budget of the universe today will be negligible, making $\chi_L$ the prime dark matter candidate.

The $\chi$ particles can scatter off each other via $Z'$ and/or $\phi'$ exchange but the cross section will be much below the values that can have observable effects \cite{Peter:2012jh}: $\sigma(\chi\chi\to \chi\chi)/m_\chi\ll 0.1~{\rm cm}^2/{\rm gr}.$ As a result, our dark matter candidate can be regarded as collision-less dark matter for structure formation and galaxy cluster collisions. For a review on the impact of self-interaction on structure formation, see \cite{Tulin:2017ara}.

Let us now discuss the impact of our model on the early universe, with special attention to the possibility of extra relativistic degrees of freedom that can affect Big Bang Nucleosynthesis (BBN) and/or CMB. The new gauge boson  ($Z'$)  and the new scalar 
($\phi$) as well as the new quarks ($T$ and $B$) are all heavier than GeV and decay long before the BBN era and the neutrino decoupling so they cannot affect these observables. The impact of $\chi_L$ and $\phi'$ on the early universe cosmology depends on their production mechanism. Within the low reheating mechanism from the beginning, their densities will be too small to be relevant. Within the freeze-out scenario,  $\chi_L$ and $\phi'$ can be produced and thermalized in the early universe via the new gauge interaction.  The freeze-out of  $\chi_L$ via $\chi_L\chi_L \to \nu_\tau \nu_\tau$ would take place at $T\sim m_\chi/20$. As long as $m_\chi/20 \gg$few MeV ({\it i.e.,} neutrino decoupling era), the annihilation products thermalize with the plasma and cannot affect BBN and CMB, either. $\phi'$ can quickly decay into the $\chi_L$ pair even before the $\chi_L$ freeze-out so it cannot affect BBN and CMB. 

Due to the $Z' $ coupling to $\nu_\tau$, the cross section of self-scattering of $\stackrel{(-)}{\nu}_\tau \stackrel{(-)}{\nu}_\tau \to  \stackrel{(-)}{\nu}_\tau \stackrel{(-)}{\nu}_\tau$  can increase. However, this enhancement cannot significantly affect the structure formation  because at the matter radiation equality era ($T\sim 2$~eV) when the structure formation becomes efficient, the rate of this scattering becomes  much lower than the Hubble rate: $n_\nu \sigma(\nu_\tau \nu_\tau \to \nu_\tau \nu_\tau)|_{T\simeq 2~{\rm eV} }\ll T^2/M_{Pl}^*|_{T\simeq 2~{\rm eV} }$ where $\sigma(\nu_\tau \nu_\tau \to \nu_\tau\nu_\tau) \sim (9 g_{Z'})^4T^2/
(4\pi m_{Z'}^4).$
%%%%%%%%%%%%%
%%%%%%%%%%
\section{Testable signals of the model\label{signal}}
%%%%%%%%%%%%%%%%
%%%%%%%%%%%%%
In the previous section, we found that in our model the axial NSI coupling, $\epsilon_{\tau\tau}^{Au}=\epsilon_{\tau\tau}^{Ad}$, can be as large as  $\mathcal{O}(1)$ which as shown in \cite{Abbaslu:2023vqk} will have observable impact on the number of neutral current events at the far detector of DUNE. This coupling can also affect the diffusion of the tau neutrinos out of the supernova core \cite{Cerdeno:2023kqo,Fiorillo:2023ytr,Cerdeno:2021cdz}. 
In fact, the main goal of the present model is to obtain axial NSI with large couplings. To build a viable anomaly-free gauge interaction that leads to such NSI, we introduced a new gauge boson, $Z'$, heavy quarks ($T$ and $B$) and the dark matter candidate ($\chi_L$). An exciting aspect of our model is that each of these particles may show up in the upcoming experiments, making our toy model ``super-testable." Let us discuss their potential signatures one by one. 

\textit{Discovery of $Z'$:} A low mass gauge boson such as $Z'$ coupled to $u$ and $d$ can be discovered as a resonance in the dijet distribution in the proton proton colliders. In fact, the LHC data with an integrated luminosity of 35.9 fb$^{-1}$  has already set a bound of $\sim 0.2$ on the coupling of $Z'$ with a mass of $\sim 30$~GeV to the quarks \cite{CMS:2019xai}. The high luminosity phase of the LHC with 3 ab$^{-1}$ integrated luminosity (100 times more data) is expected to probe couplings one order of magnitude lower, delving into the most interesting range of the parameter space of this model.
However, this discovery cannot determine whether the $Z'$ coupling to the quarks is axial or of vector-type.  Observing a deviation of the number of the neutral current events at the far detector of DUNE, along with the stringent bounds on $\epsilon^{Vq}$ from the other neutrino experiments, can testify the axial form of the interaction.

\textit{Signatures of $\chi_L$:} As we saw in the previous section, this model naturally offers a candidate for a dark matter which can undergo spin dependent dark matter scattering off nuclei. Direct dark matter search experiments already constrain the dark matter mass to the values lighter than $O(1)$~GeV within this model with $\epsilon^{Au}=\epsilon^{Ad}\sim 1$. The upcoming direct dark matter searches may find $\chi_L$.  For the detection of a GeV scale DM, the main obstacle is lowering the detection energy threshold, $E_{th}$. The minimum velocity that a DM particle should have in order to lead to an observable recoil energy is $v_{min}\simeq \sqrt{2 E_{th}/m_{DM}}$ \cite{Cirelli:2024ssz}. The number of DM scattering events will be given by $\sigma_{SD}^N\times\int_{v_{min}}^{v_{esc}} \exp (-v^2/v_0^2) v^2 dv$ where $v_0\simeq 250$ km/sec and  $v_{esc}\simeq 544$ km/sec \cite{Cirelli:2024ssz}. As seen from Fig.~\ref{fig:both_plots}, 
for $\epsilon_{\tau \tau}^{Au}=\epsilon_{\tau \tau}^{Ad}=0.9$ ({\it i.e.,} $\lambda_T=\lambda_B=4$), the PICO experiment limits $m_\chi<2$ GeV. This experiment has an energy threshold of 2.45 keV \cite{PICO:2019vsc}. If $E_{th}$ is lowered to 1.2 keV, with the same exposure,  the number of events for $m_{DM}=1$ GeV will be the same as the number of events for  $m_{DM}=2$ GeV and $E_{th}=2.45$ keV. That is $m_{DM}$ down to 1~GeV can be probed, covering the most interesting part of the parameter space.
%Th)e indirect dark matter search bounds from dwarf galaxies are satisfied in our model. If the production mechanism for $\chi_L$
%in the early universe is freeze-out with annihilation to dark sector, we would expect fractional contribution to extra relativistic degrees of freedom in the early universe which can be eventually tested by improvements on the BBN and CMB measurements.

\textit{Discovery of $T$ and $B$:} As emphasized in the previous section
$t_R$, $c_R$, $b_R$ and $s_R$ with $U(1)$ charge of 1 are taken to be mass eigenstates. However, mass eigenstates $\tilde{u}_R$
and $\tilde{T}_R$ ($\tilde{d}_R$
and $\tilde{B}_R$) are a mixture of $u_R$ and $T_R$ ($d_R$ and $B_R$)
 with opposite $U(1)$ charges.  The same consideration applies to left-handed components of these quarks, $\tilde{u}_L$
and $\tilde{T}_L$ ($\tilde{d}_L$
and $\tilde{B}_L$). In fact, in the case of left-handed components, we should also consider that while $u_L$ and $d_L$ form a $SU(2)$ doublet, $T_L$ and $B_L$ do not. Putting these considerations together, we find out that the flavor changing neutral currents with vertices including third and/or fourth generations are forbidden, implying that Kaons, $D$-mesons and $B$-mesons will behave as the SM predicts. However, we can have vertices such as 
\begin{equation} \bar{\tilde{B}}_L\gamma^\mu \tilde{d}_L Z'_\mu \ , \ \bar{\tilde{B}}_L\gamma^\mu \tilde{d}_L Z_\mu\ , \  \bar{\tilde{T}}_L\gamma^\mu \tilde{u}_L Z'_\mu \ {\rm and} \  \bar{\tilde{T}}_L\gamma^\mu \tilde{u}_L Z_\mu\end{equation}
given by the relevant gauge coupling times the corresponding mixing angle. Similarly, we have the following vertices
$$ \bar{\tilde{B}}_R\gamma^\mu \tilde{d}_R Z'_\mu \ {\rm and} \  \bar{\tilde{T}}_R\gamma^\mu \tilde{u}_R Z'_\mu \ .$$ Remembering that $u_L$ and $d_L$ are defined to be perpendicular to mass eigenstates $\tilde{c}_R$, $\tilde{t}_R$, $\tilde{s}_R$ and $\tilde{b}_R$, we can therefore write the charged current interactions as
\begin{equation} W_\mu (\bar{u}_L \ \bar{\tilde{c}}_L \ \bar{\tilde{t}}_L) \gamma^\mu V_{CKM} \left( \begin{matrix} d_L \cr \tilde{s}_L \cr \tilde{b}_L\end{matrix}\right)+H.c.=\end{equation} $$ W_\mu (\bar{\tilde{u}}_L\cos \alpha + \bar{\tilde{T}}_L\sin \alpha\ \ \bar{\tilde{c}}_L \ \bar{\tilde{t}}_L) \gamma^\mu V_{CKM} \left( \begin{matrix} \tilde{d}_L\cos\beta+\tilde{B}_L\sin\beta \cr \tilde{s}_L \cr \tilde{b}_L\end{matrix}\right)+H.c.$$
where $\sin \alpha \simeq \lambda_{Tu} \langle H\rangle /M_T$ and 
$\sin \beta \simeq \lambda_{Bd} \langle H\rangle /M_B$.  As mentioned in sect.~\ref{model}, the dominant decay modes of $\tilde{B}$ and $\tilde{T}$ are $\tilde{B}\to W\tilde{u}$, $\tilde{B}\to Z\tilde{d}$, $\tilde{T}\to W\tilde{d}$ and $\tilde{T}\to Z\tilde{u}$.
Remember that  $\epsilon^{Au}_{\tau\tau}=\epsilon^{Ad}_{\tau\tau}$ is suppressed by the inverse of  $\langle \phi\rangle^2=M_T^2/\lambda_T^2= M_B^2/\lambda_B^2$. Thus, to have sizable NSI, the masses of $T $ and $B$ should be close to the present lower bounds. $B$ and $T$ can therefore be pair produced via QCD interaction at the LHC and detected as jet+$W$ or jet+$Z$. With 140~fb$^{-1}$, the ATLAS collaboration has ruled out $T$ and $B$ quarks of 1530 GeV mass \cite{ATLAS:2024zlo}. That is the ratio of signal to root of background is equal or less than 2 ({\it i.e.,} $\mathcal{S}/\sqrt{\mathcal{B}}\leq 2$). If the masses of these particles are close to 1500 GeV, this means that with 140 $fb^{-1}$ integrated luminosity,  the ratio  $\mathcal{S}/\sqrt{\mathcal{B}}\simeq 2$. At the high luminosity phase of LHC with 3 ab$^{-1}$, both the signal and the background will be larger by a factor of $3 ~{\rm ab^{-1}}/140 ~{\rm ab^{-1}}=3000/140$. Thus, the ratio $\mathcal{S}/\sqrt{\mathcal{B}}$ can become as large as  $\sqrt{3000/140}\times 2\sim 9.2$ which means there will be a good chance of $T$ and $B$ discovery at more than $5 \sigma$ C.L.

Notice that the unitarity of the $3\times 3$ light quark mixing will be violated by $\cos\alpha \cos\beta$. Taking $\lambda_{Tu}\langle H\rangle \sim M_{Tu} \sim \sqrt{m_uM_T}$
and $\lambda_{Bd}\langle H\rangle \sim M_{Bd} \sim \sqrt{m_dM_B}$, we find $1-\cos \alpha \cos \beta \sim 3 \times 10^{-6}$ which is too small to be tested by the unitarity of the CKM matrix.

%%%%%%%%%%%%%%
%%%%%%%%%%%%
\section{Possible generalization of the toy model\label{gener}}
%%%%%%%%%%%%%%%
%%%%%%%%%%%%%%
In sect. \ref{model}, we presented a model for $\epsilon_{\tau \tau}^{Au}=\epsilon^{Ad}_{\tau \tau}$. The reason why we have focused on the $\tau \tau$ component is that this component is less experimentally restricted. We could also assign a $U(1)$ charge to the second generation of leptons and obtain $\epsilon_{\mu \mu}^{Au}=\epsilon^{Ad}_{\mu \mu}\ne 0$. As long as the $U(1)$ charges of the second and third generation leptons summed up to 9, the $U(1)-SU(2)-SU(2)$ anomaly would be canceled. Exploring the parameter space is beyond the scope of this paper. Similarly to the models for vector NSI presented in \cite{Farzan:2019xor,Farzan:2016wym}, arbitrary flavor structure can be obtained by adding sterile neutrinos with a $U(1)$ charge mixed with the active neutrinos.

To cancel the $U(1)-SU(3)-SU(3)$ anomaly, we have added a pair of quarks. The lower bound on the masses of these colored particles leads to an upper bound on $\epsilon^{Au}=\epsilon^{Ad}$. Instead of introducing new quarks, we could assign $U(1)$ charges to $s_R$ and $c_R$, $\alpha_s$ and $\alpha_c$, such that $\alpha_s+\alpha_c=6$. This requires new Higgs doublets with suitable $U(1)$ charges to give mass to the first and second quarks. Then, to cancel various $U(1)$ and $U_Y(1)$ anomalies, new charged particles (charged both under $U_{em}(1)$ and new $U(1)$) should be added. In the minimal version, with the addition of only a pair of chiral fermions, $E_L$ and $E_R$, with an electric charge of 1 and different $U(1)$ charges, the anomaly cancellation requires fractional $U(1)$ charges.  

The mass of $E$ should be obtained after breaking of the $U(1)$ symmetry via a Yukawa coupling of form $\phi \bar{E}_L E_R$. The new fermion is colorless so the lower bounds on its mass are less stringent. As a result, $\langle \phi \rangle$ can be smaller than the case with colorful new fermions. Thus, $\epsilon^{Au}=\epsilon^{Ad}$ can be larger in this version of the model without new particles.

In sect. \ref{model}, we imposed a $Z_2$ symmetry to make $\chi_L$ stable and a suitable dark matter candidate. If future direct dark matter searches  do not find it
(or more precisely, for a  given value of $\epsilon^{Au}_{\tau \tau}\sim 1$, the direct dark matter searches restrict the $\chi$ mass to below 100~MeV), we may relax the $Z_2$ symmetry and allow its decay. The dark matter should then find another candidate.

%%%%%%%%%%%%%%%%%%%%%
%%%%%%%%%%%%%%%%%%%%%
\section{Summary and discussion\label{summary}}
%%%%%%%%%%%%%%%%%%%%%%%
%%%%%%%%%%%%%%%%%%%%%%
We have introduced a $U(1)$ gauge model that leads to axial non-standard interactions between the $u$ and $d$ quarks and $\nu_\tau$. The model respects the isospin symmetry, predicting equal couplings for the $u$ and $d$ quarks, $\epsilon^{Au}_{\tau \tau
}=\epsilon^{Ad}_{\tau \tau
}$ with values of order of 1 which can be tested by the DUNE experiment \cite{Abbaslu:2023vqk}. In order for the effective four Fermi formalism to be viable for describing the scattering of the neutrino beam at DUNE (with energies of $\sim$GeV) off nuclei, the mass of the gauge boson should be larger than $\sim 10$ GeV.  Such a new gauge boson can be discovered by high luminosity LHC as a resonance of dijets.
Discovery of $Z'$ at the LHC together with the deviation of the observed number of neutral current events from the SM predictions at the far detector of DUNE  would establish the axial nature of $Z'$ coupling to the quarks.

To obtain axial coupling, the $U(1)$ charges of right-handed and left-handed first generation quarks have to be opposite which leads to a $U(1)-SU(3)-SU(3)$ anomaly. To cancel the anomaly, we introduce new quarks with masses of $\sim 1500$ GeV.  In our model, the mass generation mechanism for the light quarks is not Yukawa coupling to the SM Higgs but mixings with new heavy quarks. In other words, there will be a seesaw mechanism, explaining the smallness of the $u$ and $d$ quark masses relative to the top mass. The heavy quarks can be pair produced at the hadron colliders and can then decay to the first generation quarks plus a $W$ or $Z$ gauge boson. Being only slightly heavier than the present lower bound, there is a bright prospect of discovery of the heavy quarks at the near future by the LHC. 

Finally, to cancel the $U(1)$ anomaly, a SM singlet chiral fermion, $\chi_L$, has to be added. Imposing a $Z_2$ symmetry, $\chi_L$ can be made stable and therefore a suitable dark matter particle. The dark matter can undergo spin dependent scattering off quarks. The bounds from spin dependent direct dark matter searches already push the $\chi_L$ mass to be smaller than 2~GeV. The future spin dependent dark matter experiments can discover $\chi_L$. We have discussed the $\chi_L$ production mechanism in the early universe and pointed out three alternative scenarios to obtain the observed dark matter abundance. In our model, the indirect dark matter searches are expected to yield null results and the dark matter will act as collision-less at structure formation as well as at collision of the galaxy clusters.

In summary, our model is super-testable with four potential discoveries around the corner: (1) The impact of $\epsilon_{\tau \tau}^{Au}=\epsilon_{\tau \tau}^{Ad}$ on the number of neutral current events at the far detector of DUNE; (2) discovery of $Z'$ at the high luminosity LHC; (3) discovery of heavy quarks at the high luminosity LHC and (4) discovery of $\sim$ GeV mass dark matter at spin dependent direct dark matter search experiments.

We have discussed a few possible generalizations of the model. In particular, we have discussed how an arbitrary lepton flavor structure for $\epsilon^{Au}=\epsilon^{Ad}$ can be obtained. We have also discussed a variation of the model without heavy quarks but with new charged singlets.

\section*{Acknowledgements} 
This work is based on research funded by Iran National Science Foundation
(INSF) under project No.4031487.
This work has partially  been supported by the European
Union$'$s Framework Programme for Research and Innovation Horizon 2020 under grant H2020-MSCA-ITN2019/860881-HIDDeN as well as under the Marie
Sklodowska-Curie Staff Exchange grant agreement No
5
101086085-ASYMMETRY. YF would like to acknowledge support from ICTP through the Associates
Programme and from the Simons Foundation through
grant number 284558FY19.
YF thanks the ICTP high energy group for their hospitality during her visit when this paper was written.
\bibliography{biblio}

%\begin{thebibliography}{10}

%\end{thebibliography}

\end{document}